\documentclass[amsmath,amssymb,aps,prl,superscriptaddress,twocolumn,floatfix]{revtex4-2}
\usepackage{graphicx} 
\usepackage{wrapfig}
\usepackage{bm}
\usepackage{xcolor}
\usepackage{MnSymbol}
\usepackage{hyperref} 
\usepackage{comment}

\usepackage{subfigure}

\newcommand{\xCornell}{Department of Physics, Cornell University, Ithaca, NY, USA}
\newcommand{\xEwha}{Department of Physics, Ewha Womans University, Seoul, South Korea}

\begin{document}

\title{Switching between superconductivity and current density waves in Bernal bilayer graphene}
\author{Jun Ho Son}
\affiliation{\xCornell}
\author{Yi-Ting Hsu}
\affiliation{Department of Physics, University of Notre Dame, Notre Dame, Indiana 46556, USA}
\author{Eun-Ah Kim}
\affiliation{\xCornell}
\affiliation{\xEwha}

\begin{abstract}
An out-of-plane magnetic field can always suppress superconductivity. 
In Bernal-stacked bilayer graphene (BBG), recently observed activation of superconductivity (SC) through either in-plane magnetic fields or proximate spin-orbit coupling (SOC) offers a rare instance of switching superconductivity {\emph on}. 
To understand this, we must first examine the non-superconducting state. We propose an incommensurate current density wave (CrDW) driven by van Hove singularities away from the zone corners as a competing order. 
We note that the two switches, the in-plane field and the SOC, both break spin degeneracy. 
Our parquet renormalization group analysis reveals that breaking spin degeneracy shifts the balance from CrDW, favored under spin degeneracy, to SC when degeneracy is lifted. Driven by purely repulsive interactions, the pairing symmetry of the resulting SC is $p/d$-wave.
The presence of CrDW accounts for the non-linear $I-V$ behavior in the normal state and suggests potential anomalous Hall effects due to time-reversal symmetry breaking. We further predict that reducing screening could enhance SC.
\end{abstract}

\maketitle
\begin{figure*}
    \centering
    \includegraphics[width=16cm]{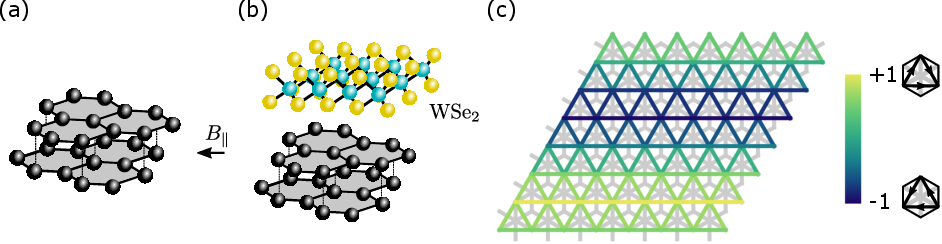}
    \caption{(a) Illustration of the first way of activating SC in BBG: Applying in-plame magnetic field (b) Illustration of the second way of activating SC in BBG: Stacking a monolayer WSe$_2$. (c) The real space picture of the CrDW phase we propose to emerge in spin-degenerate BBG instead of SC. It is manifest as as incommensurate modulation of the time-reversal odd bond operators. Since the majority of the weight for the hole band of our interest is on a single sublattice of one layer, we illustrated triangular configuration of nearest-neighbor bonds between such sites. The insets next to the legend shows which direction of the current corresponds to the plus/minus signs in the color map.} 
    \label{fig:fig1}
\end{figure*}

 The rich complexity of phase diagrams resulting from competition between superconductivity (SC) and other competing orders is the hallmark of strongly correlated superconductivity. 
Most common experimental knobs control the itineracy of charge carriers by either changing the carrier density through doping or changing bandwidth through pressure. 
 These controls enhance or suppress superconductivity in an analog fashion. Recent experiments on Bernal-stacked bilayer graphene (BBG) presented an unprecedented scenario of digital control: in-plane magnetic field $B_{||}$ \cite{Zhou2022} (Fig.~\ref{fig:fig1}(a)) and proximate spin-orbit coupling through interfacing with WSe$_2$ \cite{Zhang2023,Holleis2023} (Fig.~\ref{fig:fig1}(b))  each popped superconducting phases into already rich phase diagram \cite{DelaBarrera2022,Seiler2022,Seiler2023}. Despite clear differences between the two control switches and observed differences in the normal (off-state) state phase diagrams, the striking similarity calls for a theoretical approach that transcends the differences. 

 Earlier theoretical efforts mostly approached the phenomena from the analog control perspective, discussing how in-plane field \cite{Szabo2022, Wagner2023,Curtis2023,Shavit2023} or proximity to WSe$_2$ \cite{Chou2022_2, JimenoPozo2023,Curtis2023,Shavit2023} will increase  $T_{c}$ in already superconducting system. However, this is at odds with the fact that BBG does not superconduct without $B_{||}$ or WSe$_2$ proximate spin-orbit coupling. In fact, the regime of the phase space  that $B_{||}$ or spin-orbit coupling flips on superconductivity exhibits high electrical resistivity with a non-linear current($I$)-voltage($V$) relations \cite{Zhou2022} without $B_{||}$ or spin-orbit coupling \footnote{The proximate spin-orbit coupling is turned off when the vertical displacement field direction pushes electrons away from WSe$_2$.}.
\textcite{Dong2023_1} proposed a scenario in which $B_{\parallel}$ can activate superconductivity, assuming superconductivity to be mediated by the quantum critical fluctuation of putative valley-polarization quantum critical point. However, the proximate Ising spin-orbit coupling will explicitly break valley symmetry in spin-valley locked manner in BBG-WSe$_2$. 

 Here, we consider an under-appreciated possibility of density waves. Most of the literature on symmetry breaking in BBG focused on ferromagnetism of various incarnations such as isospin ferromagnetism\cite{Chichinadze2022, Dong_isospinf, Szabo2022, Xie2023, Zhang2023, Shavit2023} or inter-valley coherent state \cite{Xie2023,Zhang2023}. However, such ferromagnets will yield metallic states. Whereas the observed non-linear $I$-$V$ curve is reminiscent of that in incommensurate charge density waves 
\cite{Fleming1979,Lee1979, Gruner1988}. Through our theoretical analysis using the renormalization group (RG) method, we propose an incommensurate, intra-valley current density wave (CrDW) order which connects two spots on the Fermi surface within the same valley to compete with SC in BBG closely. In the real space picture, the CrDW state will show an incommensurate modulation of time-reversal odd bond operators, or \textit{current operators} (see Fig.~\ref{fig:fig1}(c)). Our proposed CrDW state, due to different signs of order parameters in the two valleys, spontaneously breaks time-reversal symmetry.

\begin{figure*}
    \centering
    \includegraphics[width=16cm]{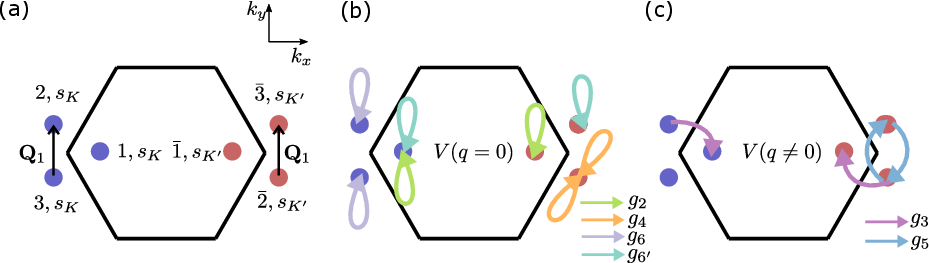}
    \caption{ 
    (a) Schematics for our six-patch models, where the blue patches ($n=1,2,3$) are from valley $K$ with spin $s_K$, and the red patches ($n=\bar{1},\bar{2},\bar{3}$) are from valley $K'$ with spin $s_{K'}$. The arrow indicates one of three CrDW ordering wave vectors, $\mathbf{Q_1},\mathbf{Q_2},\mathbf{Q_3}$. The hexagon labels the Brillouin zone. (b) Schematics of all the considered $q=0$ part of the interactions in our six-patch model. Each four-fermion interaction is sketched as a pair of fermion hopping processes, where the arrow points from an annihilation to a creation operator. (c) Similar schematics to (b) for $q\neq 0$ part of the interactions}
    \label{fig:fig2}
\end{figure*}

 The key observation behind our theoretical analysis is that van Hove singularities (vHS) near the Fermi level $E_F$ in BBG (see Fig. \ref{fig:fig2}(a)) create ``hot spots'' slightly away from the zone corners, which can promote various ordering possibilities. In particular, our proposed CrDW order connects two hot spots within the same valley and has three possible wavevectors $\mathbf{Q}_{1}$, $\mathbf{Q}_{2}$, and $\mathbf{Q}_{3}$ with an estimated magnitude $|\mathbf{Q}_{i}| \approx 1.28 \times 10^{-2} \AA^{-1}$; see Fig.~\ref{fig:fig2}(a) for illustration of $\mathbf{Q}_{1}$. We simplify the complex Fermi surface to be represented by the six momentum patches centered at the hot spots associated with vHS (see Fig. \ref{fig:fig2}(a), (b)) and employ a parquet RG analysis \cite{Schulz1987,Furukawa1998}  to explore competition among superconductivity, density waves, isospin ferromagnetism, and nematic phases to test if the presence or absence of spin-degeneracy have the potential of switching between different competing states in a manner that transcends microscopic differences between BBG with $B_{||}\neq0$ and BBG-WSe$_2$.

\emph{Model---} 
 When only the nearest-neighbor hoppings within each layer and the nearest-neighbor interlayer couplings, the band structure of BBG is famously known to have a quadratic dispersion. Incorporating non-nearest neighbor hoppings leads to so-called trigonal warping in which the quadratic band bottom broadens into a flat region featuring three minima related to each other through $C_{3}$ symmetry. Because of this effect, at the low hole-doping, the Fermi surface features three hole pockets near each valley, but upon increasing the doping level, they undergo Lifshitz transition and combine into a single hole pocket. The vHS of our interest is associated with this Lifshitz transition.

 The vHS is slightly away from the zone corners in BBG. Hence, three vHS exist per valley near the Fermi level $E_F$, as depicted by red spots and blue spots in Fig.~\ref{fig:fig2}(a)  (see SM Section I for more detail). We describe spin-degenerate BBG using this six-patch model with spin-degenerate patches $s=\uparrow,\downarrow$, and describe BBG-WSe$_2$ and BBG under an in-plane magnetic field $B_{\parallel}$ with a six-patch model with spin-polarized patches $s'=\bar{s}$ and  $s=s'$, respectively. In the spin-degenerate and spin-split cases, the number of fermion flavor on each patch is therefore $N_f=2$ and $N_f=1$, respectively. 
Most recent parquet RG analysis for various (quasi) two-dimensional materials focused on cases with number of fermion flavors $N_{f} \geq 2$ \cite{Isobe2018,Lin2019, Hsu2020, Lin2020,Park2021,Wu2023,You2022,Lu2022}. In order to elucidate the role of spin degeneracy, we consider repulsive screened Coulomb interaction well-suited to describe BBG and compare and contrast the RG flow between $N_f=2$ and $N_f=1$, keeping all else fixed. 

 The number of fermion flavor $N_f$ can affect the allowed inter- and intra-patch interactions through symmetries and fermionic statistics \cite{Hsu2021}. For spin-degenerate BBG with $N_f=2$, the interactions that obey time-reversal $\mathcal{T}$ and point group $D_{3}$ symmetries are given by (see Fig. \ref{fig:fig2}(b)) 
 \begin{equation}
\begin{split}
\label{eq:fourfermionint}
& H_{\text{int}} = \frac{1}{2\nu_{0}} \sum_{s,s'} \sum_{n=1}^{3} \sum_{m \neq n}  g_{2} \psi^{\dagger}_{s,n} \psi_{s,n} \psi^{\dagger}_{s',\bar{n}} \psi_{s',\bar{n}} \\
& \quad \quad \quad \quad   + 2 g_{3} \psi^{\dagger}_{s,n} \psi_{s,m} \psi^{\dagger}_{s',\bar{n}} \psi_{s',\bar{m}} + 2 g_{6'} \psi^{\dagger}_{s,n} \psi_{s,n}\psi^{\dagger}_{s',\bar{m}} \psi_{s',\bar{m}}  \\
& \quad \quad \quad \quad  + \frac{g_{4}}{2} \left[  \psi^{\dagger}_{s,n} \psi_{s,n} \psi^{\dagger}_{s',n} \psi_{s',n} + \psi^{\dagger}_{s,\bar{n}} \psi_{s,\bar{n}} \psi^{\dagger}_{s',\bar{n}} \psi_{s',\bar{n}} \right] \\
& \quad \quad \quad \quad + g_{5} \left[  \psi^{\dagger}_{s,n} \psi_{s,m} \psi^{\dagger}_{s',m} \psi_{s',n} + \psi^{\dagger}_{s,\bar{n}} \psi_{s,\bar{m}} \psi^{\dagger}_{s',\bar{m}} \psi_{s',\bar{n}} \right]\\
& \quad \quad \quad \quad  + g_{6} \left[ \psi^{\dagger}_{s,n} \psi_{s,n} \psi^{\dagger}_{s',m} \psi_{s',m} + \psi^{\dagger}_{s,\bar{n}} \psi_{s,\bar{n}} \psi^{\dagger}_{s',\bar{m}} \psi_{s',\bar{m}} \right], 
\end{split}
\end{equation}
where $\psi^{\dagger}_{s,n}$ creates an electron with spin $s=\uparrow,\downarrow$ from patch $n$, and the subscript for intra-patch momentum $\textbf{k}$ is suppressed, where momentum conservation is assumed. In Eq.~\eqref{eq:fourfermionint}, $\nu_{0}$ is a constant 
that appears in the ultra-violet (UV) divergence of the bare Cooper susceptibility $\Pi_{\text{pp}} \sim \frac{1}{2} \nu_{0} \log \frac{\Lambda}{T} $ at finite temperature $T$, where $\Lambda$ is the UV cutoff set by the patch size; it is introduced to make $g_{i}$'s dimensionless. $g_4$ term is the intra-patch density-density interaction, $g_2$, $g_6$, and $g_{6'}$ are the inter-patch density-density interactions, and $g_3$, $g_5$ are intra-valley scatterings. Of these, $g_{2}$, $g_{4}$, $g_{6}$, and $g_{6'}$
originate from $q=0$ part of the screened Coulomb interactions $V(q)$. We set their values to be identical and denote the bare value by $V(q=0)$  (see Fig.~\ref{fig:fig2}(b)). On the other hand, the intra-valley scatterings  $g_{3}$ and $g_{5}$ originate from $V(q\neq 0)$ part of the screened Coulomb interaction. We assign the identical bare values $V(q \neq 0)$ to $g_{3}$ and $g_{5}$ as well(see Fig.~\ref{fig:fig2}(c)).
Note that Umklapp scattering is not allowed since the vHS are not on Brillouin zone boundaries. Moreover, we expect that inter-valley scatterings are negligible due to the large momentum transfers, while small inter-valley scatterings do not qualitatively change our results.
For $N_f=1$ systems with spin-polarized patches, it was shown in Ref. \onlinecite{Hsu2021} that only four interactions in $H_{\text{int}}$ survive and remain independent. Specifically, the intra-patch density-density interaction $g_{4}$ vanishes due to the Fermi exclusion principle. Moreover, $g_{5}$ and $g_{6}$ become indistinguishable. Therefore, we consider only $g_{2}$,$g_{3}$, $g_{5}$, and $g_{6'}$ interactions in Eq.~\eqref{eq:fourfermionint} for BBG-WSe$_2$ or BBG under an in-plane field $B_{\parallel}$ with the bare interaction strength $V(q=0)$ for $g_{2}$ and $g_{6'}$ and $V(q\neq 0)$ for $g_{3}$; $V(q\neq 0)- V(q=0)$ for $g_{5}$.

\emph{Parquet RG Approach---} 
We now perform an RG analysis to identify the relevant instability in our six-patch models for BBG. 
Specifically, under the RG scheme in which we progressively integrate out the modes with energy between energy $E$ and $E+dE$ at each step, we track how interaction $g_{i}$'s and susceptibility $\chi_{P}$ for each possible instability $P$ associated with an order parameter $\Delta_{P;ss';nm} \Psi^{\dagger}_{s,n}\Psi_{s',m}^{(\dagger)}$ are renormalized under the RG flow at one-loop level. Specifically, we solve the RG equations for different choices of initial values $(V(q=0),V(q \neq 0) )$ and extract the exponent $\beta_{P}$ that characterize the asymptotic power-law divergence of $\chi_{P}$. The order $P$ with the most strongly divergent $\chi_{P}$ is identified with the dominant instabilities.
Such RG analyses have been previously done for the $N_f=2$ cases\cite{Isobe2018,Lin2019,Hsu2020} and $N_f=1$ \cite{Hsu2021} cases separately for different physical systems. Here we explore purely repulsive interactions with the bare values of the interaction organized for BBG, and compare and contrast the difference the lifting of spin degeneracy makes by comparing $N_f=2$  and $N_f=1$ on equal footing. 
We briefly review the key aspects of the RG calculation for completeness, leaving the explicit form of equations and derivations in the supplementary materials (SM Section II). 

 The first step of our calculation is numerically solving one-loop RG equations for $g_{i}$'s as functions of the RG scale $y = \frac{1}{2} \log^{2} \frac{\Lambda}{E}$, where $\Lambda$ is the UV cutoff set by the patch size. Each RG flow $g_{i}(y)$ asymptotically approach the form $G_{i}/(y_{c}-y)$ near $y = y_{c}$ at which a set of $g_{i}$'s diverge. We extract $y_{c}$ and $G_{i}$'s that characterize the asymptotic behavior of the RG flow from the numerical solutions and use these parameters to determine how each susceptibility $\chi_{P}$ diverges near $y=y_{c}$. In particular, using the asymptotic form of $g_{i}(y)$, we show in SM that $\chi_{P} $ diverges as $\chi_P(y)\propto (y_c-y)^{2\beta_P+1}$ with the exponent $\beta_{P}$. The most negative $\beta_{P}$ determines the dominant channel of instability $P$.

Of all competing instabilities we considered (see SM section III for the full list of vHS-driven spin, charge, and pairing instabilities and the corresponding $\beta_{P}$'s) we found two instabilities that dominate and closely compete in addition to the much-considered valley ferromagnetism are:
\begin{equation}
\label{eq:beta}
\beta_{\text{$p/d$-SC}} = G_{2} - G_{3},
\end{equation}
for the $p/d$-wave superconductivity (SC)
\footnote{reduced symmetry upon spin-degeneracy lifting mixes $p$ and $d$ wave} and
\begin{equation}
\label{eq:beta2}
\beta_{\text{CrDW}} = d_{3}(y_{c}) \left[ -N_{f} G_{3}+N_{f} G_{5} -  (N_{f} -1)G_{6} \right],
\end{equation}
for the CrDW. In Eq.~\eqref{eq:beta2}, $d_{3}(y) = \frac{\gamma_{3}}{\sqrt{\gamma_{3}^{2} + 2y}}$ parameterizes the degree of particle-hole nesting between two patches $\textit{not}$ related to each other by $\mathcal{T}$, and $\gamma_{3} \gg 1 $ is directly related to UV divergence of the corresponding bare particle-hole susceptibility $\Pi_{\text{ph}} \sim \gamma_{3} \nu_{0} \log \frac{\Lambda}{T} $ at finite temperature $T$.
 A current density wave ordering was first considered as a candidate description of the pseudogap state in high $T_c$ cuprates \cite{ddw4}. Although there is no definitive evidence of current density wave in cuprates, a commensurate current density wave state have been proposed \footnote{\cite{Feng2021,Wilson2023} and references therein} as an explanation of observed time-reversal symmetry breaking in Kagome metals \cite{Wilson2023}.  

\begin{figure*}
    \centering
    \includegraphics[width=16cm]{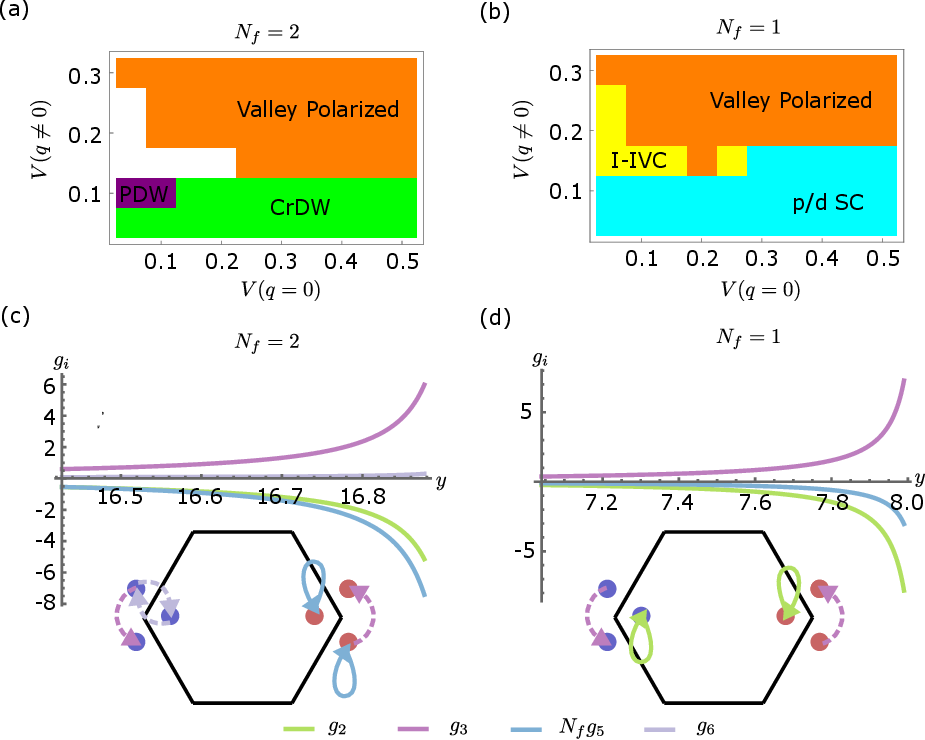}
    \caption{(a) The phase diagram for spin-degenerate case $N_f=2$.  (b)  The phase diagram with spin-degeneracy lifted $N_f=1$. Both phase diagrams are obtained at a strong nesting degree $\gamma_3=8$ with different choices of initial interaction strengths $(V(q=0),V(q \neq 0))$. Purple and yellow squares represent pair-density wave states and incommensurate versions of inter-valley coherent (IVC) states in which the nesting vector for the charge density wave connects vHS in different valleys. The white square represents a region in which the numerical solutions $g_i(y)$ do not fit well into the asymptotic form used, and the phase cannot be determined with the method given here. (c) The RG flows for the spin-degenerate case $N_{f}=2$ of inter- and intra-patch interactions $g_i(y)$ near $y_{c}$, obtained numerically by solving the RG equations at $(V(q=0),V(q \neq 0)) = (0.2,0.05)$,  $\gamma_{3}=8$. The arrows on the hexagon illustrate the driving interaction channels for the CrDW, the leading instability for the choice of parameter used here, with the attractive channels illustrated as solid lines and repulsive channels as dashed lines.  (d)  The RG flows of $g_i(y)$'s for the spin-split $N_{f}=1$, obtained with the same set of $\gamma_{3}$ and $(V(q=0),V(q \neq 0))$ as (c). The inset with the hexagon shows the driving interaction channels for the SC, with solid and dashed lines having the same meaning as (c). $g_{6}$ is identical to $g_{5}$ in this case and hence omitted. }
    \label{fig:fig3}
\end{figure*}

 The phase diagram in Fig.~\ref{fig:fig3} shows the dominant instabilities as a function of bare coupling strengths of $q=0$ and $q\neq0$ part of the interaction, setting the parameter that controls the particle-hole nesting strength $\gamma_{3} = 8$. Irrespective of spin degeneracy, the valley ferromagnetism dominates in the limit of large $V(q \neq 0)$. In this limit,  strongly repulsive $g_{5}$, which is a finite-$q$ repulsive interaction between two valley densities, stabilizes the valley ferromagnet via Stoner-like mechanism. However, when the interaction is less screened (smaller $V(q \neq 0)$) CrDW dominates when spin degeneracy is not split (Fig.~\ref{fig:fig3}(a)). Upon breaking the spin degeneracy, the CrDW is replaced by $p/d$-wave superconductivity. 

 We can understand the role of spin-degeneracy in controlling the balance between the CrDW and SC by observing how the factor $d_3$ and the linear combination of $G_i$'s contribute to the magnitude of the exponent $\beta_{CrDW}$ in Eq.~\eqref{eq:beta2}. The typical value of $d_3(y)<1$ reflects the subdominance of the log-divergent particle-hole nesting driving CrDW next to the log$^2$-divergence of the Cooper instability. However, spin-degeneracy amplifies the interactions in the particle-hole channel. Hence, with a reasonable degree of nesting (i.e., large $\gamma_3$) the CrDW exponent dominates with spin degeneracy ($N_f=2$). 
  An inspection of the RG flow of $g_2$, $g_3$, and $g_4$, which are divergent near $y=y_c$ and contribute to the susceptibility exponents related to CrDW and SC confirms this understanding. Fig.~\ref{fig:fig3}(c-d) shows that the signs of all the running couplings will indeed cause the susceptibilities for both instabilities to diverge (i.e., negative exponent in Eq.~\eqref{eq:beta} and Eq.~\eqref{eq:beta2}) irrespective of the spin-degeneracy. However, the factor of fermion flavor number $N_f$ that only weighs $g_5$ decisively makes the exponent for CrDW more negative in the spin-degenerate system, masking superconducting instability. Lifting the spin-degeneracy, either through an in-plane field or a proximate spin-orbit coupling will reveal the superconducting instability.  

\emph{Conclusion---} Based on a parquet RG analysis for spin-degenerate and spin-split Bernal bilayer graphene (BBG), we propose that the observed high-resistivity state in spin-degenerate BBG is a CrDW state driven by the vHS near the Fermi level. Lifting spin degeneracy via $B_{\parallel}$ or SOC from the WSe$_2$ layer promotes SC. 
The proposed mechanism can be tested in future experiments in several ways. (1) The proposed CrDW with wavevector $\mathbf{Q}_{1}$ would couple to charge density wave at $2\mathbf{Q}_{1}$ (see SM Section V for the detailed discussion). Given the length scale of the charge density wave $\lambda_{\text{CDW}} \sim 12 \text{nm}$ similar to twisted moire length scale, the CDW can be detected through scanning tunneling microscopy \cite{TBG}. 
(2) The time-reversal symmetry breaking can be detected through anomalous Hall effect \cite{Yang2020}.
(3) Reducing screening will promote superconductivity. A nontrivial way to change screening is by varying the spin-degeneracy splitting. With stronger spin-splitting, the system will have lower hole density and weaker screening. Our mechanism predicts stronger spin-splitting to promote superconductivity. This scenario is consistent with the fact that 
SC appears at lower hole densities as one increases $B_{\parallel}$ \cite{Zhou2022}. Intriguingly, in Ising SOC with the estimated strength $\lambda_{I} \approx 0.7 \text{meV}$ in BBG-WSe$_2$ in Ref.~\onlinecite{Zhang2023}
provides a stronger source of spin-splitting than $B_{\parallel}$  with the Zeeman energy $E_{Z} < 0.1 \text{meV}$ in Ref.~\onlinecite{Zhou2022}. In fact, in the phase diagram for BBG-WSe$_2$, SC starts to show up at a lower hole density with much higher $T_{c}$ compared to the $B_{\parallel}$-driven SC in \cite{Zhou2022}. Controlling the strength of Ising SOC through the relative angle between BBG and WSe$_2$ will be an interesting test \footnote{S. Nadj-Perge, private communication}. 
In closing, we remark on the $C_{3}$ breaking observed in  Refs.~\cite{Zhou2022,Zhang2023}. 
Nematic order is subdominant to both CrDW and SC (see SM Section III) within parquet RG framework. However, nematic order may appear as a vestigial order \cite{Fernandes2019} of charge density wave coupled to CrDW.

\textit{Acknowledgement---} J.H.S and E.-A.K acknowledge funding from AFOSR MURI grant No. FA9550-21-1-0429. Y.-T.H. acknowledges support from NSF Grant No. DMR-2238748. The authors acknowledge 
helpful discussions with Philip Kim, Stevan Nadj-Perge, Jason Alicea, Long Ju, Andrea Young, and Andrey Chubukov. 

\bibliography{bibliography}
\end{document}


\title{Supplementary Materials for Switching between superconductivity and current density waves in Bernal bilayer graphene}
\author{Jun Ho Son}
\affiliation{\xCornell}
\author{Yi-Ting Hsu}
\affiliation{Department of Physics, University of Notre Dame, Notre Dame, Indiana 46556, USA}
\author{Eun-Ah Kim}
\affiliation{\xCornell}
\affiliation{\xEwha}
\maketitle

\section{Band structure and van Hove singularities of the Bernal bilayer graphene under high displacement field}

 Here, we provide a more detailed discussion on the band structure and van Hove singularities in Bernal bilayer graphene in the regime of our interest and show that van Hove singularities are arranged in the pattern illustrated in the Fig.~2 of the main text. For concreteness, we focus on the band structure near the $K$ valley, while the band structure near the other valley can be straightforwardly obtained from the time-reversal action. We also discuss the effect of the spin-degeneracy lifting perturbations.

 \begin{figure}
     \centering
     \includegraphics[width=16.5cm]{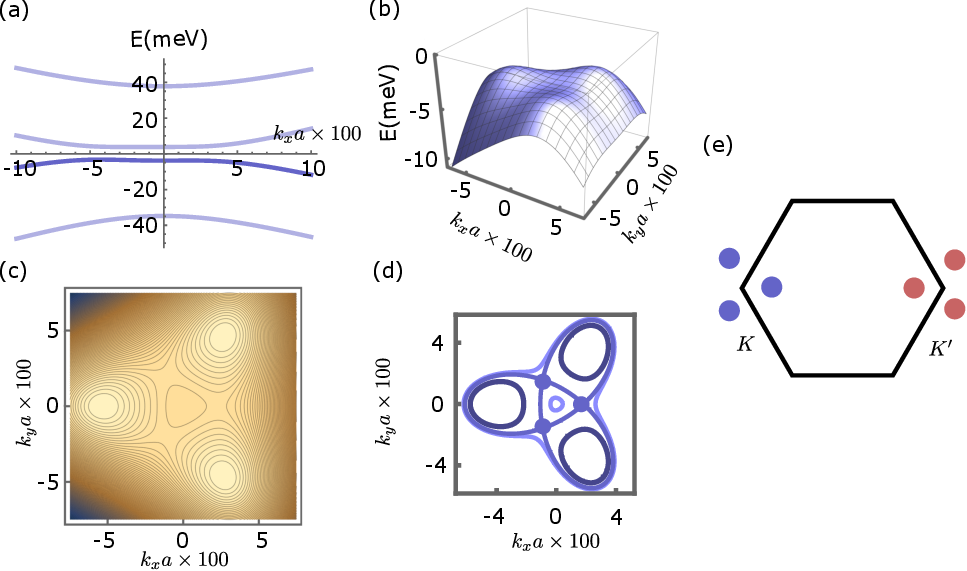}
     \caption{(a) The band dispersion of the BBG near $K$ valley along the line cut $k_{y}=0$. The hole band of our interest is emphasized with a darker color. (b) The dispersion of the hole band of our interest, plotted in 3D. (c) Equal energy contours of the hole band of our interest, with color map in which darker color represents contours with energy farther away from $E=0$. (d) Three selected equal energy contours at $\mu=-36.0$meV (slightly below the vHS filling), $\mu= -37.3$ meV (right at the vH filling), and $\mu = =-38.3$meV  (slightly above the vH filling), with darker colors representing higher fillings. Three vHS at $\mu= -37.3$ meV are marked as dots as well. (e) configuration of vHS in the Brillouin zone.}
     \label{fig:band}
 \end{figure}

 Following the discussion in the supplementary material of \cite{Zhang2023}, we take a look at the band structure near $K$-valley with the following $k$-space Hamiltonian:
\begin{equation}
\label{eq:klinearham}
\begin{split}
    &  H = \mathbf{\Psi}_{\mathbf{k}}^{\dagger} \begin{pmatrix}
        \frac{U}{2} & \frac{\sqrt{3}}{2a} \gamma_{0} (k_{x} + i k_{y}) & -\frac{\sqrt{3}}{2a} \gamma_{4} (k_{x} + i k_{y}) & - \frac{\sqrt{3}}{2a} \gamma_{3} (k_{x} - i k_{y}) \\
        \frac{\sqrt{3}}{2a} \gamma_{0} (k_{x} - i k_{y}) & \frac{U}{2} + \Delta & \gamma_{1} &  - \frac{\sqrt{3}}{2a} \gamma_{4} (k_{x} - i k_{y}) \\
        -\frac{\sqrt{3}}{2a} \gamma_{4} (k_{x} - i k_{y}) & \gamma_{1} & -\frac{U}{2} + \Delta &  \frac{\sqrt{3}}{2a} \gamma_{0} (k_{x} + i k_{y}) \\
        - \frac{\sqrt{3}}{2a} \gamma_{3} (k_{x} + i k_{y}) & -\frac{\sqrt{3}}{2a} \gamma_{4} (k_{x} - i k_{y})  & \frac{\sqrt{3}}{2a} \gamma_{0} (k_{x} - i k_{y}) & -\frac{U}{2}
    \end{pmatrix} \mathbf{\Psi}_{\mathbf{k}} \\
    & \mathbf{\Psi}_{\mathbf{k}}^{\dagger}
     = \begin{pmatrix} \psi^{\dagger}_{\mathbf{k},A, 1} & \psi^{\dagger}_{\mathbf{k},B, 1} & \psi^{\dagger}_{\mathbf{k},A, 2} & \psi^{\dagger}_{\mathbf{k},B, 2}
    \end{pmatrix}
\end{split}
    \end{equation}
 In the above equation, $\psi^{\dagger}_{\mathbf{k}, s=A,B, l=1,2}$ is an electron creation operator residing in the sublattice $s$ and layer $l$ with lattice momentum $\mathbf{k}$ measured from $K$ point, and $a$ is the lattice constant of graphene. The above Hamiltonian is the most general form of the Hamiltonian near $K$-valley up to linear order in $\mathbf{k}$. Since we are interested in a regime with a small doping level, neglecting $O(k^{2})$ pieces may be justified.
 
 For the parameters $(\gamma_{0},\gamma_{1},\gamma_{3}, \gamma_{4}, \Delta)$ we use the widely quoted values $\gamma_{0} = 2.61$eV, $\gamma_{1}=0.361$eV, $\gamma_{3}=0.283$eV, $\gamma_{4}=0.138$eV, $\Delta= 0.015$eV obtained from fitting the ab-initio band structure for BBG without displacement field \cite{Jung2014}. For $U$ which parameterizes the inter-layer potential difference from displacement field, following \cite{Zhang2023}, we assume $ U = d_{\perp} D/\epsilon_{\text{BBG}}$ between two layers with $d_{\perp}=0.33\text{nm}$ (distance between two layers in BBG) and $\epsilon_{\text{BBG}} \approx 4.3$ (dielectric constant of BBG), and the displacement field $D=1$V/\text{nm} at which the competition between SC and other states we are interested in are manifest in the experiment.

 As a first step, we look at the dispersion of the four bands near $K$ valley obtained from the Hamiltonian in Eq.~\eqref{eq:klinearham}. In Fig.~\ref{fig:band}(a), we plot the dispersion along the line cut with $k_y = 0$. We see that the lowermost band and the uppermost band are well-separated from two bands near $E=0$. These relatively large separations come from strong inter-layer tunneling between $B$ sublattice of layer 1 and $A$ sublattice of layer 2  which are on top of each other in the Bernal stacking. This tunneling is represented in the parameter $\gamma_{1}$ in the Hamiltonian Eq.~\eqref{eq:klinearham}. Hence, in these two bands, majority of wavefunctions are concentrated on these two sites.
 
 Meanwhile, the other two bands have the majority of weights on the $A$ sublattice of layer 1 and $B$ sublattice of layer 2 respectively. In particular, strong displacement field $D$ effectively polarizes the layer index in these two bands as well. Among these two bands, the hole band is the one relevant for the experiments we would like to account for in BBG. 

 Now, we take a closer look at the dispersion and the Fermi surface topology of this hole band. Fig.~\ref{fig:band}(b) shows the dispersion in the 3D plot. One can see that the band bottom is not quadratic; trigonal warping which originates from non-nearest neighbor hoppings``broadens" the quadratic band bottom in the simplified nearest-neighbor tight-binding model for BBG. 

 It is instructive to look at the equal-energy contours, plotted in Fig.~\ref{fig:band}(c) and (d), to connect the trigonal warping effect and the emergence of vHS. As one can see in Fig.~\ref{fig:band}(c), at small chemical potential, the band features three small hole pockets. As one increases the filling, these three hole pockets merge into a large hole pocket and a small electron pocket at the center which disappear quickly as one increases the filling. One can clearly see the origin of the Lifshitz transition in our description, and the trio of vHS near $K$ valley is associated with this Lifshitz transition. 
 
 In Fig.~\ref{fig:band}(d), we plotted the shape of the Fermi surfaces slightly below the van Hove fillings, precisely at the van Hove fillings, and slightly above the van Hove fillings. One can see the merging of three hole pockets at the van Hove filling, and the three vHs we marked as dots are precisely the points where the three hole pockets touch each other at the van Hove filling.
  
 The three van Hove singularties near each valley are related to each other through $D_{3}$ symmetry of Bernal bilayer graphene. $D_3$ symmetry does not only constrain dispersions near six van Hove singularities to be related to each other but also fix the van Hove singularities to lie on the line connecting the $\Gamma$ point and the hexagonal Brillouin zone corners -- any exceptions to this rule can only happen after going through additional Lifshitz transitions (See \cite{shtyk2017} for the example) associated with higher van Hove singularities of which the experiment shows no sign. Hence, the vHS positioned slightly away from the zone corners are expected to be general features even after taking into various effects that lead to the deviations from the ab-initio result we employ to generate the plots.

  After accounting for the trio of van Hove singularities in both valleys, one may take these vHS to be arranged in the pattern in Fig.~\ref{fig:band}(e) in the Brillouin zone. 

  Finally, we discuss the effect of spin-splitting perturbations such as the in-plane magnetic field and Ising spin-orbit coupling. Spin-degeneracy lifting splits a single van Hove filling into two, one at higher and one at lower hole doping. At the lower van Hove filling, holes with one specie of spin sits exactly at van Hove singularities, while holes with the opposite spin is below the van Hove fillings, and as illustrated in Fig.~\ref{fig:band}(d), form three Fermi pockets. As one increases the strength of spin-splitting perturbations, the three Fermi pockets at the lower van Hove fillings will be smaller. Therefore, the van Hove filling is achievable with lower hole doping.
  
\section{The full one-loop RG equations}

 For the $N_{f}=2$ case, the one-loop RG equations are given by:
\begin{equation}
    \begin{split}
        & \frac{d g_{2}}{d y} = -\left[ 1 - d_{2}(y) \right] g_{2}^{2} - 2 g_{3}^{2} + 2 d_{1}(y) \left[ - g_{2} g_{4} - 4 g_{6}' g_{6} +2 g_{6}' g_{5} \right] \\
        & \frac{d g_{3}}{d y} = - g_{3}^{2} - 2 g_{2}g_{3} + 2d_{3}(y) g_{3} \left[ g_{6}' + g_{6} - 2 g_{5} \right] \\
        & \frac{d g_{4}}{d y} = - d_{4}(y) g_{4}^{2}  + d_{1}(y)  \left[ -2 g_{2}^{2} -4 g_{6}'^{2} + g_{4}^{2} - 4 g_{6}^{2}  + 4 g_{5}g_{6}  + 2 g_{5}^{2}  \right] \\
        & \frac{d g_{5}}{d y} =  d_{3}(y) \left[ 2 g_{5}g_{6} -2 g_{5}^{2} -2g_{3}^{2} \right] -2 d_{5}(y)  g_{5}g_{6} +  d_{1}(y)  \left[ 2g_{4} g_{5} + g_{5}^{2}   \right] \\
        & \frac{d g_{6}}{d y} =  d_{3}(y) g_{6}^{2} - d_{5}(y)  \left[ g_{5}^{2} + g_{6}^{2}   \right] + 2 d_{1}(y)  \left[ g_{5} g_{6} -  g_{6}^{2} -  g_{6}'^{2} - g_{4}g_{6} + g_{4}g_{5} - 2 g_{2}g_{6}'  \right] \\
         & \frac{d g_{6}'}{d y} = - d_{5}(y) g_{6}'^{2} + d_{3}(y) \left[ g_{6}'^{2} + g_{3}^{2} \right] + 2 d_{1}(y) \left[ -2g_{6}'g_{4} + g_{2}g_{5} -2 g_{2} g_{6} + g_{5}g_{6}' - 2 g_{6}g_{6}'  \right].
    \end{split}
\end{equation}

 For the $N_{f}=1$ case, the RG equations are the following:
 \begin{equation}
    \begin{split}
        & \frac{d g_{2}}{d y} = - g_{2}^{2} - 2 g_{3}^{2} + 4d_{1}(y)   g_{6'} g_{5} +  d_{2}(y) g_{2}^{2} \\
        & \frac{d g_{3}}{d y} = - g_{3}^{2} - 2 g_{2}g_{3} + 2d_{3}(y) g_{3} \left[ g_{6}' - g_{5}  \right] \\ 
        & \frac{d g_{5}}{d y} =  d_{5}(y) g_{5}^{2}  - d_{3}(y) \left[ g_{3}^{2} + g_{5}^{2} \right] + d_{1}(y)  \left[ 2 g_{2} g_{6}'+  g_{5}^{2} +  g_{6}'^{2} \right] \\
        & \frac{d g_{6}'}{d y} = - d_{5}(y) g_{6}'^{2} + d_{3}(y) \left[ g_{6}'^{2} + g_{3}^{2} \right] - 2 d_{1}(y) \left[ g_{2} g_{5}  + g_{6}' g_{5} \right].
    \end{split}
\end{equation}
 The RG equations for $N_{f}=1$ are identical to the ones appearing in \cite{Hsu2021}. For the $N_{f}=2$ case, we set all inter-valley scatterings to zero, while the RG equations in \cite{Hsu2020} also consider non-zero inter-valley scatterings. 
  
 The key feature distinguishing the above RG equations from the usual one-loop RG equations is that each term carries a $d$ factor which accounts for UV divergence in distinct particle-particle/particle-hole channels. To see how $d$-factors are defined, we first examine the bare susceptibilities $\Pi_{ph/pp}$ in the particle-hole and particle-particle channels, which are the building blocks of the RG analysis. $\Pi_{ph/pp}$ have the form $\Pi_{\text{ph}}^{nm}(\textbf{q})= -\int d\textbf{k} \frac{f_{\textbf{k}+\textbf{q}} - f_{\textbf{k}}}{\epsilon^{n}_{\textbf{k}+\textbf{q}} - \epsilon^m_{\textbf{k}}}$ and $\Pi_{\text{pp}}^{nm}(\textbf{q})= \int d\textbf{k} \frac{1 - f_{\textbf{k}+\textbf{q}} - f_{\textbf{k}}}{\epsilon^{n}_{\textbf{k}+\textbf{q}} + \epsilon^m_{-\textbf{k}}}$, where $f_{\textbf{k}}$ is the Fermi-Dirac distribution, $\epsilon^n_{\textbf{k}}$ ($\mathbf{k}$ is measured from the center of the patch) is the patch dispersion, and the momentum transfer $\textbf{q}$ connects the relevant patches $n$ and $m$. Among all bare susceptibilities, the only two that exhibit ln$^2$ divergence as temperature $T\rightarrow 0$ are the Cooper instability $\Pi_{pp}^{n\bar{n}}(0)= \frac{\nu_{0}}{2} $ln$^2(\frac{\Lambda}{T})$ and the intra-patch particle-particle susceptibility  $\Pi_{pp}^{nn}(2\textbf{Q}_n)= \gamma_4\Pi_{pp}^{n\bar{n}}(0)$, where the UV cutoff $\Lambda$ is set by the patch size, $\nu_{0}$ is a patch dispersion-dependent parameter, $\textbf{Q}_n$ is the momentum where patch $n$ is located, and $\gamma_4$ is a parameter that captures the intra-patch nesting degree in the particle-particle channel. In contrast, the inter-patch particle-hole susceptibility $\Pi_{ph}^{nm}$ for non-perfectly nested Fermi surface and all other bare susceptibilities, such as the density of states $\Pi_{ph}^{nn}$ and the particle-particle nesting degree $\Pi_{pp}^{nm}$ for $m\neq\bar{n},n$, are expected to be at most log divergent. Therefore, we parameterize their values with respect to the density of states as $\Pi(\textbf{q})=\gamma_{\eta}\Pi^{nn}_{ph}(0)$\cite{Hsu2020,Hsu2021}, where $\eta=1,2,3,5$ corresponds respectively to $\Pi=\Pi^{nn}_{ph}$, $\Pi^{n\bar{n}}_{ph}$, $\Pi^{nm}_{ph}$, and $\Pi^{nm}_{pp}$ with $m\neq n,\bar{n}$. 

 The $d$ factor is defined to be $d_\eta(y) \equiv d\Pi(\textbf{q})/dy$ that characterizes the corresponding bare susceptibility $\Pi(\textbf{q})=\gamma_\eta\Pi_{ph}^{nn}(0)$ or $\Pi(\textbf{q})=\gamma_\eta \Pi_{pp}^{n\bar{n}}(0)$. For the ln-divergent $\Pi(\textbf{q})$, these $d$ factors decrease in $y$, and we choose to take the form $d_{\eta}(y)=\gamma_{\eta}/\sqrt{\gamma_\eta^{2} +2y}$ for $\eta=1,2,3,5$. For the ln$^2$-divergent $\Pi(\textbf{q})=\Pi_{pp}^{nn}(2\textbf{Q}_n)$, the corresponding $d$ factor is given by $d_{4}(y)=\frac{1+\gamma_{4}y}{1+y}$. From the dispersion near vHS in BBG, we expect $\gamma_{3}$ can be tuned to be large, while the other $\gamma_{\eta}$'s is close to unity.

\section{More on the test vertex method for determining the susceptibility}
  We follow \cite{Nandkishore2012,Lin2019,Hsu2020,Hsu2021} to systematically determine the dominant order. We first observe that any candidate for a possible order $\Delta_{P}$ may be identified with a linear combination of fermion bilinears $\Delta_{P;sn,s'm}\psi_{s,n}^{\dagger}\psi_{s',m}$ (for a non-superconducting order) or $\Delta_{P;sn,s'm}\psi_{s,n}^{\dagger}\psi_{s',m}^{\dagger}$ (for a superconducting order) that attains a long-range order. The susceptibility $\chi_{P}$ associated with $\Delta_{P}$ can be computed by introducing a ``test vertex" term $\delta H (\Delta_{P}, \Delta_{P}^{*})  =  \sum_{s,s'n,m,} \Delta_{P;sn,s'm}\psi_{s,n}^{\dagger}\psi_{s',m}^{(\dagger)} + \text{H.c.}$ (heretofore, we also use $\Delta_{P}$ and $\Delta_{P}^{*}$ to refer to the overall coefficient of the test vertex term) to the original Hamiltonian to $H$. The static susceptibility $\chi_{P}$ can be obtained from the following functional derivative:
  \begin{equation}
  \label{eq:chipdef}
      \exp \left[ - F[\Delta_{P}, \Delta_{P}^{*}] / T\right] = \text{Tr} \exp \left[ -\frac{H + \delta H (\Delta_{P}, \Delta_{P}^{*}) }{T}\right], \quad \chi_{P} = - T \left. \frac{\delta^{2}  F[\Delta_{P}, \Delta_{P}^{*}]}{ \delta \Delta_{P} \delta \Delta_{P}^{*}} \right\vert_{\Delta_{P} = \Delta_{P}^{*} = 0} .
  \end{equation} 
   \begin{figure}
    \centering
    \includegraphics[width=16cm]{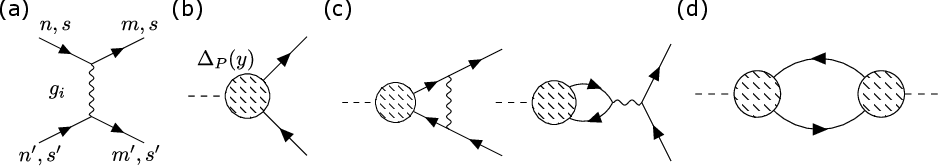}
    \caption{(a)  Feynman diagram notations for the interaction vertex $g_{i}$ (b) Feynman diagram notations for the renormalized test vertex $\Delta^{P}(y)$. (c) One-loop diagrams that contribute to the test vertex renormalization. (d) The tree-level diagram contributing to the renormalization of $\chi_{P}(y)$ governed by Eq.~\eqref{eq:chipeq}.}
    \label{fig:suppfig1}
\end{figure}
  Meanwhile, in our RG scheme in which one progressively integrates out modes occupying a thin energy shell in the partition function, the test vertex term $\delta H$ also undergoes a non-trivial renormalization when it is incorporated into the partition function and the free energy. We keep track of this renormalization to the one-loop order in $g_{i}$'s valid in the regime where the interactions can be treated perturbatively and the linear order in $\Delta_{P}$ and $\Delta_{P}^{*}$  justified from the fact that our quantity of interest $\chi_{P}$ is determined from the small-$\Delta_{P}$ behavior of $F[\Delta_{P}, \Delta_{P}^{*}]$. To these orders, in the Feynman diagram language where we use the wavy lines illustrated in Fig.~\ref{fig:suppfig1}(a) to represent the four-fermion interaction $g_{i}$'s and the shaded circle in Fig.~\ref{fig:suppfig1}(b) to represent the renormalized vertex $\Delta_{P}$, the test vertex renormalization comes from the one-loop diagrams illustrated in Fig.~\ref{fig:suppfig1}(c) upon integrating over high energy modes on the loops of the diagrams. A judicious choice of a linearly independent basis for fermion bilinears $\{ \Delta_{P} \}$ allows one to take the RG equation to the following ``diagonal" form where $d_{P} (y)=1$ or $d_{\eta}(y)$ given in the previous section and $C^{(P)}_{i}$ a numerical coefficient obtainable from the Feynman diagram combinatorics:
  \begin{equation}
  \label{eq:testvertexeq}
      \frac{d \Delta_{P}(y)}{d y} = - d_{P}(y)  \Delta_{P}(y) \sum_{i} C^{(P)}_{i} g_{i}(y).
  \end{equation}
  
  At $y \approx y_{c}$, one may take $d_{P}(y)$ and $g_{i}(y)$ to be $d_{P}(y_{c}) $ and the asymptotic solution $\frac{G_{i}}{y_{c} - y}$  respectively. Solving the differential equation with this information gives the asymptotic power law behavior for the renormalized test vertex term coefficient $\Delta_{P}(y)$, given by the following:
  \begin{equation}
  \label{eq:testvertexasymptotic}
        \Delta_{P}(y) \sim  (y_{c}-y)^{\beta_{P}}, \quad \beta_{P} =d_{P}(y_{c}) \sum_{i} C^{(p)}_{i} G_{i}.
  \end{equation}
  
  We now look at how the above asymptotic form of $\Delta_{P}(y)$ may be used to determine the asymptotic behavior of the corresponding susceptibility. For this purpose, it is useful to consider a scale-dependent susceptibility $\chi_{P}(y)$, defined identically to Eq.~\eqref{eq:chipdef} except for the test vertex term $\delta H$ which is instead taken to be the following $y$-dependent term $\delta H(y)$ ($\psi_{n,s}^{\dagger} (E)$ and $\psi_{n,s} (E)$ are fermion modes with the energy $E$, and $E(y) \geq 0$ is the energy corresponding to the RG scale $y$) for a non-superconducting order $\Delta_{P}$:
  \begin{equation}
       \delta H(y) = \sum_{n,m,s,s'} \sum_{|E| < E(y)} \Delta_{P;ns,ms'} \psi_{n,s}^{\dagger} (E) \psi_{m,s'}(E) + \text{H.c.},
  \end{equation}
  i.e., $\chi_{P}(y)$ is computed from taking expectation values only with respect to high energy fermion modes (recall that $y \sim \log^{2} \frac{\Lambda}{|E|}$, so small $y$ corresponds to high energy $E$).  $\chi_{P}(y)$ for the superconducting order can be defined similarly. By definition, $\chi_{P}(y \rightarrow \infty) = \chi_{P}$ if $\chi_{P}(y)$ is well-defined over all values of $y \geq 0$. Alternatively, $\chi_{P}(y)$ may be thought as representing the susceptibility in the presence of IR cutoffs such as temperature and external frequency/momenta.

 The idea is that $\chi_{P}(y + dy) - \chi_{P}(y)$, to the lowest non-trivial order in the interaction $g_{i}$, can be obtained from the tree-level diagram in Fig.~\ref{fig:suppfig1}(d) by using the \textit{renormalized} partition function and integrating over modes with the energy scale between $y$ and $y+dy$. Especially, the test vertices in Fig.~\ref{fig:suppfig1}(d) are the \textit{renormalized} ones represented by the shaded circles and $\Delta_{P}(y)$, and the test vertex renormalization given by Eq.~\eqref{eq:testvertexeq} encodes the lowest-order interaction effect as promised earlier.
    
  Taking this effect into account, $\chi_{P}(y)$ may be estimated from integrating over the following differential equation:
\begin{equation}
\label{eq:chipeq}
    \frac{d \chi_{P}(y)}{ dy} = d_{P}(y) \left\vert \frac{\Delta_{P}(y)}{\Delta_{P}(0)} \right\vert^2,
\end{equation}
  with the factor $\left\vert \frac{\Delta_{P}(y)}{\Delta_{P}(0)} \right\vert^2$ on the right side coming from the fact that we are using the renormalized test vertices. Taking $\Delta_{P}(y)$ to be the asymptotic form Eq.~\eqref{eq:testvertexasymptotic} near $y_{c}$ and $d_{P}(y)$ to be a fixed value $d_{P}(y_{c})$ as before gives the asymptotic power law divergence of $\chi_{P}(y) \sim (y_{c}-y)^{ 2\beta_{P} + 1} $ near $y = y_{c}$. Hence, we related the text vertex renormalization governed by the RG equation Eq.~\eqref{eq:testvertexeq} and the asymptotic solution Eq.~\eqref{eq:testvertexasymptotic} to the IR behavior of $\chi_{P}(y)$. Especially, if $\beta_{P} < -\frac{1}{2}$ and $\alpha_{P} < 0$, $\chi_{P}(y)$ is IR-divergent near $y=y_{c}$, signaling the potential onset of the order $\Delta_{P}$ at the scale above $y_{c}$

  In the table~\ref{table:nf2} and ~\ref{table:nf1}, we give a full list of fermion bilinears $\{ \Delta_{P} \}$ which lead to the diagonal RG equations for $\Delta_{P}(y)$, the orders they correspond to, and the formula for the $\beta_{P}$, the power law exponent associated with late-time behaviors of $\chi_{P}(y)$. One can see from the tables that we indeed treat superconductivity, pair/spin/charge density waves, spin/valley polarized phases, and nematic orders on equal footings. Especially, going beyond the standard ``fast-parquet" approach and incorporating all $\log$-divergences allow one to treat spin/valley polarized and nematic phases that are not traditionally considered in the parquet RG framework. The CrDW state in the main text corresponds to CDW$_{b-}$ and DW$_{b-}$ in the table, and the PDW$_a$ and DW$_{c+}$ are the pair density wave state and the inter-valley density wave state that occupy a small region in the phase diagram of the main text.

\begin{table}
  \begin{tabular}{|c|c|c|}
  \hline
  Name  & Fermion bilinears & $\beta_{P}$ \\
  \hline
  $s/f$-wave SC & $\displaystyle \sum_{n=1}^{3} \psi^{\dagger}_{n,s} \sigma^{\mu}_{ss'} \psi^{\dagger}_{\bar{n},s'}$ & $G_{2} + 2G_{3}$   \\
 \hline
  $p/d$-wave SC & $\displaystyle \sum_{n=1}^{3} \alpha_{n} \psi^{\dagger}_{n,s} \sigma^{\mu}_{ss'} \psi^{\dagger}_{\bar{n},s'}, \, \sum_{n=1}^{3} \alpha_{n} = 0  $ & $G_{2} - G_{3}$ \\
  \hline
  PDW$_{a}$ & $ \psi^{\dagger}_{n,s} \sigma^{y}_{ss'} \psi^{\dagger}_{n,s'}$ & $d_{4}(y_{c})  G_{4}$ \\
  \hline
  PDW$_{b\pm}$ & $ \psi^{\dagger}_{n,s} \sigma^{y}_{ss'} \psi^{\dagger}_{m,s'} (+)/ \psi^{\dagger}_{n,s} \sigma^{\mu \neq y}_{ss'} \psi^{\dagger}_{m,s'} (-) $ & $d_{5}(y_{c}) \left[G_{6} \pm G_{5} \right]$   \\
  \hline
  PDW$_{c}$ &$ \psi^{\dagger}_{n,s} \sigma^{\mu}_{ss'} \psi^{\dagger}_{\bar{m},s'}$  & $d_{4}(y_{c}) G_{6'} $ \\
  \hline 
  SDW/CDW$_{a}$ &$ \psi^{\dagger}_{n,s} \sigma^{\mu}_{ss'} \psi_{\bar{n},s'}$  & $ -d_{2}(y_{c}) G_{2} $  \\
  \hline 
  CDW$_{b\pm}$ &$ \psi^{\dagger}_{n,s}\psi_{m,s} \pm \psi^{\dagger}_{\bar{n},s}\psi_{\bar{m},s} $  & $d_{3}(y_{c}) \left[-G_{6} + 2 G_{5} \pm 2 G_{3} \right]$ \\
  \hline 
  SDW$_{b\pm}$ & $ \psi^{\dagger}_{n,s} \sigma^{\mu\neq 0}_{ss'} \psi_{m,s'} \pm \psi^{\dagger}_{\bar{n},s} \sigma^{\mu\neq 0}_{ss'} \psi_{\bar{m},s'} $ & $-d_{3}(y_{c}) G_{6} $  \\
  \hline
  SDW/CDW$_{c \pm }$ & $ \psi^{\dagger}_{n,s} \sigma^{\mu}_{ss'} \psi_{\bar{m},s'} \pm \psi^{\dagger}_{\bar{n},s} \sigma^{\mu}_{ss'} \psi_{m,s'} $  & $-d_{3}(y_{c}) \left[G_{6'} \pm G_{3} \right]$ \\
  \hline 
  Spin-polarized  & $ \displaystyle \sum_{n=1}^{3} \psi^{\dagger}_{n,s} \sigma^{\mu \neq 0}_{ss'} \psi_{n,s'} \pm \psi^{\dagger}_{\bar{n},s} \sigma^{\nu \neq 0 }_{ss'} \psi_{\bar{n},s'} $  & $-d_{1}(y_{c}) \left[G_{4} + 2 G_{5} \right]$  \\
  \hline 
  Valley-polarized  & $ \displaystyle \sum_{n=1}^{3} \psi^{\dagger}_{n,s}\psi_{n,s} - \psi^{\dagger}_{\bar{n},s} \psi_{\bar{n},s} $  & $-d_{1}(y_{c}) \left[2G_{2} + 4G_{6'} - G_{4} -4G_{6} + 2 G_{5}  \right]$  \\
  \hline 
  Nematic$_{+}$  & $ \displaystyle \sum_{n=1}^{3} \alpha_{n} \left[ \psi^{\dagger}_{n,s}\psi_{n,s} + \psi^{\dagger}_{\bar{n},s} \psi_{\bar{n},s} \right], \, \sum_{n=1}^{3} \alpha_{n} = 0  $   &  $-d_{1}(y_{c}) \left[-2G_{2} - 2G_{6'} - G_{4} +2G_{6} -  G_{5}  \right]$  \\
  \hline 
  Nematic$_{-}$  & $ \displaystyle \sum_{n=1}^{3} \alpha_{n} \left[ \psi^{\dagger}_{n,s}\psi_{n,s} - \psi^{\dagger}_{\bar{n},s} \psi_{\bar{n},s} \right], \, \sum_{n=1}^{3} \alpha_{n} = 0  $   &  $-d_{1}(y_{c}) \left[2G_{2} + 2G_{6'} - G_{4} +2G_{6} -  G_{5}  \right]$ \\
  \hline 
  Nematic$_{\text{spin}}$ & $ \displaystyle \sum_{n=1}^{3} \alpha_{n} \left[ \psi^{\dagger}_{n,s} \sigma^{\mu\neq 0}_{ss'} \psi_{n,s'} \pm \psi^{\dagger}_{\bar{n},s} \sigma^{\nu\neq 0}_{ss'} \psi_{\bar{n},s'} \right], \, \sum_{n=1}^{3} \alpha_{n} = 0  $  & $-d_{1}(y_{c}) \left[G_{4}  -  G_{5}  \right]$   \\
  \hline
\end{tabular}
\caption{List of possible instabilities considered for the $N_{f}=2$ model. $\sigma^{\mu=0,x,y,z}_{ss'}$ is the Pauli matrix, and spin indices $s$ and $s'$ are implicitly summed over. The Greek alphabet superscripts for the Pauli matrices are reserved for the case in which they can be any of the four matrices. $n$ and $m$ are patch indices that can be $1$,$2$,$3$ $\bar{1}$, $\bar{2}$, or $\bar{3}$ with $n\neq m$ unless stated otherwise. $\bar{n}$ maps $n=1,2,3$ to $\bar{1},\bar{2},\bar{3}$ and vice versa.}
\label{table:nf2}
\end{table}

\begin{table}
  \begin{tabular}{|c|c|c|}
  \hline
  Name  & Fermion bilinears & $\beta_{P}$ \\
  \hline
  $s/f$-wave SC & $\displaystyle \sum_{n=1}^{3} \psi^{\dagger}_{n} \psi^{\dagger}_{\bar{n}}$ & $G_{2} + 2G_{3}$   \\
 \hline
  $p/d$-wave SC & $\displaystyle \sum_{n=1}^{3} \alpha_{n} \psi^{\dagger}_{n} \psi^{\dagger}_{\bar{n}}, \, \sum_{n=1}^{3} \alpha_{n} = 0  $ & $G_{2} - G_{3}$ \\
  \hline
  PDW$_{b}$ & $ -\psi^{\dagger}_{n}\psi^{\dagger}_{m} $ & $d_{4}(y_{c}) G_{5} $   \\
  \hline
  PDW$_{c}$ &$ \psi^{\dagger}_{n}  \psi^{\dagger}_{\bar{m}}$  & $d_{4}(y_{c}) G_{6'}$ \\
  \hline 
  DW$_{a}$ &$ \psi^{\dagger}_{n} \psi_{\bar{n}}$  & $ -d_{5}(y_{c}) G_{2} $  \\
  \hline 
  DW$_{b\pm}$ &$ \psi^{\dagger}_{n}\psi_{m} \pm \psi^{\dagger}_{\bar{n}}\psi_{\bar{m}} $  & $d_{3}(y_{c}) \left[G_{5} \pm  G_{3} \right]$ \\
  \hline
  DW$_{c \pm }$ & $ \psi^{\dagger}_{n} \psi_{\bar{m}} \pm \psi^{\dagger}_{\bar{n}} \psi_{m} $  & $-d_{3}(y_{c}) \left[G_{6'} \pm G_{3} \right]$ \\
  \hline 
  Valley-polarized  & $ \displaystyle \sum_{n=1}^{3} \psi^{\dagger}_{n}\psi_{n} - \psi^{\dagger}_{\bar{n}} \psi_{\bar{n}} $  & $-d_{1}(y_{c}) \left[G_{2} + 2G_{6'} + 2G_{5} \right]$  \\
  \hline 
  Nematic$_{+}$  & $ \displaystyle \sum_{n=1}^{3} \alpha_{n} \left[ \psi^{\dagger}_{n}\psi_{n} + \psi^{\dagger}_{\bar{n}} \psi_{\bar{n}} \right], \, \sum_{n=1}^{3} \alpha_{n} = 0  $   & $-d_{1}(y_{c}) \left[-G_{2} + G_{6'} - G_{5}  \right]$   \\
  \hline 
  Nematic$_{-}$  & $ \displaystyle \sum_{n=1}^{3} \alpha_{n} \left[ \psi^{\dagger}_{n}\psi_{n} - \psi^{\dagger}_{\bar{n}} \psi_{\bar{n}} \right], \, \sum_{n=1}^{3} \alpha_{n} = 0  $   & $-d_{1}(y_{c}) \left[G_{2} - G_{6'} - G_{6}  \right]$ \\ 
  \hline
\end{tabular}
\caption{List of possible instabilities considered for the $N_{f}=1$ model. As before, $n$ and $m$ are patch indices that can be $1$,$2$,$3$ $\bar{1}$, $\bar{2}$, or $\bar{3}$ with $n\neq m$ unless stated otherwise. $\bar{n}$ maps $n=1,2,3$ to $\bar{1},\bar{2},\bar{3}$ and vice versa. There is no flavor index in this case.
}
\label{table:nf1}
\end{table}

\section{The phase diagram for the small $\gamma_{3}$ }

  \begin{figure}
    \centering
    \includegraphics[width=16cm]{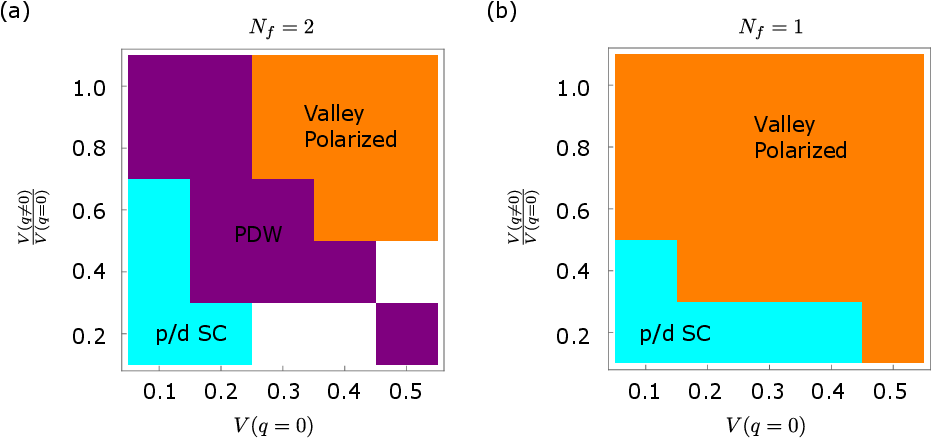}
    \caption{(a) The phase diagram at $\gamma_{3}=2$, $N_{f}=2$, for different choices of $V(q=0)$ and $V(q\neq 0)$ (b) The phase diagram obtained from the same set of parameters as (a) except $N_{f}=1$}
    \label{fig:suppfig2}
\end{figure}

 As we saw in the main text, the experimental results on (SC) in the Bernal bilayer graphene (BBG) are mostly consistent with the regime with large $\gamma_{3}$. However, it is interesting to take a look at the regime with small $\gamma_{3}$, which may be achievable in the experiments through applying stronger displacement field to BBG. Fig.~\ref{fig:suppfig2} is the phase diagrams we obtained from different choices of $(V(q),V(q\neq 0),N_{f})$ when $\gamma_{3}$ is small ($\gamma_{3}=2$), using the methods described in the main text and the previous two sections of the supplementary materials.

  As one can see in Fig.~\ref{fig:suppfig2}(a), when $N_{f}=2$ and $\gamma_{3}$ is small, the CrDW phase in the main text is suppressed and is replaced by the SC phase.  Additionally, the PDW phase occupies a much larger region in the phase diagram. 

  The SC phase in Figure.~\ref{fig:suppfig2}(a) has the same SC order parameter structure as the SC in the $N_{f}=1$ phase diagram in the main text, but due to the spin rotation $\text{SU}(2)_{K} \times \text{SU}(2)_{K'}$ symmetries for each valley, spin-singlet and spin-triplet pairings are degenerate. In the putative experiments, we expect that the degeneracy is lifted through weak inter-valley Hund's coupling that breaks $\text{SU}(2)_{K} \times \text{SU}(2)_{K'}$ into a single $\text{SU}(2)$. Alternatively, in 2D, thermal fluctuations suppress long-range orders associated with $\text{SU}(2)$ symmetry breaking in two spatial dimensions at any finite temperature \cite{Friedan1980, Chakravarty1989}. Hence, if inter-valley Hund's couplings are sufficiently small, these thermal fluctuations instead completely destroy any long-range orders associated with the usual SCs. However, charge-$4e$ Cooper pairs that are singlet under $\text{SU}(2)_{K} \times \text{SU}(2)_{K'}$ may still retain the (quasi-)long-range order, leading to the charge-$4e$ SC. This scenario is also discussed in \cite{Curtis2023}. This opens up a possibility for realizing charge-$4e$ SCs in spin-degenerate BBG. 
  
  Meanwhile, the phase diagram in Fig.~\ref{fig:suppfig2}(b) for the $N_{f}=1$ case with $\gamma_{3}=2$ is largely similar to the $\gamma_{3}=8$ case in the main text except for the suppression of the IVC phase.

  We close this section with two remarks: First, the charge-$4e$ SC we mentioned can also arise as a vestigial order of the PDW phase as well \cite{Berg2009, Fernandes2019}, purveying a unifying perspective of the $N_{f}=2$ phase diagram. Second, while the PDW phase in itself is not consistent with neither non-linear charge transport nor high resistivity observed for the phase competing with SCs in BBG, one can access a CDW phase from the PDW phase by condensing a certain vortex \cite{Berg2009}. This CDW phase is also incommensurate because the Cooper pairs in the parent PDW phase have momenta incommensurate with the underlying lattice but is distinct from any of the density wave phases in the main text. Hence, it may serve as an alternative candidate for the phase competing with the SC. However, this CDW phase requires a rather ad-hoc, delicate mechanism with a dubious microscopic origin, and we propose that the CrDW phase in the main text provides a more straightforward explanation for the competing order.

\section{Charge density wave induced from the current density wave}

 In this section, we discuss at the level of the Landau-Ginzburg functionals how charge density waves can be induced from the CrDW order, following the approach developed in the context of the interplay between spin density waves and charge density waves in cuprates \cite{Zachar1998}. The order parameters for the Landau-Ginzburg functionals of our interest are complex order parameters ($\Phi_{\mathbf{Q}_{1}}$, $\Phi_{\mathbf{Q}_{2}}$, $\Phi_{\mathbf{Q}_{3}}$) reserved for the CrDWs and $(\rho_{2\mathbf{Q}_{1}}, \rho_{2\mathbf{Q}_{2}}, \rho_{2\mathbf{Q}_{3}})$  for the charge density waves. $\mathbf{Q}_{1}$, $\mathbf{Q}_{2}$ and $\mathbf{Q}_{3}$ in the subscripts label three possible CrDW wavevectors illustrated as arrows in the Fig. 2(a) of the main text; the charge density wave order parameters have wavevectors twice of those of the CrDW order parameters. We define $\mathbf{Q}_{1}$, $\mathbf{Q}_{2}$ and $\mathbf{Q}_{3}$ so that the counterclockwise rotation $R_{2\pi/3}$ and the reflection $\mathcal{R}_{y}$ with respect to the $y$-axis implement the following transformations:
 \begin{equation}
 \label{eq:sym1}
     R_{2\pi/3}\mathbf{Q}_{1} = \mathbf{Q}_{2}, \, R_{2\pi/3} \mathbf{Q}_{2} = \mathbf{Q}_{3}, \, R_{2\pi/3} \mathbf{Q}_{3} = \mathbf{Q}_{1},
 \end{equation}
 \begin{equation}
 \label{eq:sym2}
     \mathcal{R}_{y}\mathbf{Q}_{1} = \mathbf{Q}_{1}, \, \mathcal{R}_{y}\mathbf{Q}_{2} = \mathbf{Q}_{3}, \, \mathcal{R}_{y}\mathbf{Q}_{3} = \mathbf{Q}_{2}. 
 \end{equation}
 The CrDW order parameters for $-\mathbf{Q}_{1}$, $-\mathbf{Q}_{2}$, and $-\mathbf{Q}_{3}$ are simply given by complex conjugates, i.e., $\Phi_{-\mathbf{Q}_{i}} = \Phi_{\mathbf{Q}_{i}}^{*}$ for $i=1,2,3$, and are not considered separately. Similarly, $\rho_{-2\mathbf{Q}_{i}} = \rho_{2 \mathbf{Q}_{i}}^{*}$. 

 We will consider the following Landau-Ginzburg potential:
\begin{equation}
\label{eq:LG}
    F(\{ \Phi_{\mathbf{Q}_{i}}\}, \{ \rho_{2\mathbf{Q}_{i}}\}) = \frac{r}{2}\sum_{i} |\rho_{2\mathbf{Q}_{i}}|^{2} + F_{\text{CrDW}}(\{ \Phi_{\mathbf{Q}_{i}}\}) + \sum_{i} \lambda \Phi_{\mathbf{Q}_{i}}^{2} \rho_{2 \mathbf{Q}_{i}}^{*} + \text{c.c.}
\end{equation}
First two terms are Landau-Ginzburg potentials for  $\rho_{2\mathbf{Q}_{i}}$'s and $\Phi_{\mathbf{Q}_{i}}$'s and indepedently. Because we are interested in the scenario with no primary vHS-driven instability towards charge-density wave orders, we set $r>0$ at the first term and drop cubic or higher-order terms in $\rho_{2\mathbf{Q}_{i}}$'s. The precise form of $F_{\text{CrDW}}(\{ \Phi_{\mathbf{Q}_{i}}\})$ in the second term is not important for our discussion, but the parameters in $F_{\text{CrDW}}(\{ \Phi_{\mathbf{Q}_{i}}\})$ are tuned so that $F_{\text{CrDW}}(\{ \Phi_{\mathbf{Q}_{i}}\})$ lead to at least one non-zero order parameter at the minima to account for our setups in which there is a vHS-driven instability towards the CrDW. The final term describes the lowest-order coupling between $\Phi_{\mathbf{Q}_{i}}$'s and $\rho_{2\mathbf{Q}_{i}}$'s allowed by the symmetries of BBG with the complex coefficient $\lambda$: the $C_{3}$ rotation, the reflection $\mathcal{R}_{y}$, time-reversal $\mathcal{T}$, and the lattice translation symmetry. The $C_{3}$ rotation simply changes the subscripts of order parameters acccording to Eq.~\eqref{eq:sym1}. However, order parameters transform more non-trivially under $\mathcal{R}_{y}$ and $\mathcal{T}$ because these two symmetries switch $K$ and $K'$ valley. $\mathcal{R}_{y}$ ($\mathcal{T}$) map order parameters $\Phi_{\mathbf{Q}_{i}}$ to $-\Phi_{\mathcal{R}_{y}\mathbf{Q}_{i}}$ ($-\Phi_{\mathbf{Q}_{i}}^{*}$) and $\rho_{2\mathbf{Q}_{i}}$ to $\rho_{2\mathcal{R}_{y}\mathbf{Q}_{i}}$ ($\rho_{2\mathbf{Q}_{i}}^{*}$); the minus sign for the symmetry actions for $\Phi_{\mathbf{Q}_{i}}$'s reflects the fact that the CrDW order parameters have opposite signs between two valleys. Finally, the lattice translation symmetry requires each term to be a product of order parameters whose wavevectors sum up to zero. $\mathcal{T}$ requires charge density wave order parameters to be coupled to \textit{square} of the CrDW order parameters. Hence, the lowest-order coupling is cubic. The lattice translation symmetry requires the wavevector of the charge density wave order parameters to be twice of the CrDW order parameters. This requirement naturally explains our choice of wavevectors for charge-density waves. 

 Now, we discuss the effect of small but non-zero $\lambda$. First, we assumed that at least one CrDW order parameter is non-zero at the minima of $F_{\text{CrDW}}(\{ \Phi_{\mathbf{Q}_{i}}\})$. Small $\lambda$ in $F(\{ \Phi_{\mathbf{Q}_{i}}\}, \{ \rho_{2\mathbf{Q}_{i}}\})$ will shift the location of the minima quantitatively from that of $F_{\text{CrDW}}(\{ \Phi_{\mathbf{Q}_{i}}\})$, but if $\lambda$ is sufficiently small, at the minima, there will be still non-zero CrDW order parameter. Meanwhile, $\lambda$ has a much more dramatic effect on $\rho_{2\mathbf{Q}_{i}}$. From the condition $\frac{\partial F(\{ \Phi_{\mathbf{Q}_{i}}\}, \{ \rho_{2\mathbf{Q}_{i}}\}) }{\partial \rho_{2\mathbf{Q}_{i}}^{*}} = 0$ at the minima, one can see that there will be a charge density wave order parameter $\rho_{2\mathbf{Q}_{i}}$ that acquires a non-zero expectation value $\frac{\lambda}{r} \Phi_{-\mathbf{Q}_{i}}^{2}$. Hence, our analysis using the Landau-Ginzburg potential given in Eq.~\eqref{eq:LG} serves as a proof-of-concept that the CrDW instabilities can induce charge-density wave orders whose wavevector is twice that of the CrDW. Incorporating higher-order terms in Eq.~\eqref{eq:LG} will not modify our conclusion. 

  We close this section with the quantitative estimate of the charge density wave periodicity. Recall that $\mathbf{q}_{i}$'s connect two different van Hove singularities within the same valley. For the $k$-space Hamiltonian in Eq.~\eqref{eq:klinearham} with parameters given in SM Section I., $|\mathbf{Q}_{i}| = 0.0128 \AA^{-1}$. $2\mathbf{Q}_{i}$ wavevector of the charge density wave then gives the charge density wave periodicity $\lambda_{\text{CDW}} \approx 12.3 \text{nm}$. The real-space charge modulation has a very long periodicity spanning $\sim 50$ unit cells, but $\lambda_{\text{CDW}}$ is still much smaller than the typical sample size utilized for the experiments (a few microns by a few microns) and should be visible in the real-space imaging technique.

 \bibliography{bibliography}